\documentclass[prl,superscriptaddress,twocolumn,showpacs,amsmath,amssymb,floatfix]{revtex4}
\usepackage{epsfig}

\newcommand{\cd}[2]{\begin{array}{c} #1 \\ #2 \end{array}}
\newcommand{\eq}[1]{(\ref{#1})}

\newcommand{\dd}{{\mathrm d}}

\newcommand{\Z}{{\mathbb Z}}

\newcommand{\Kanazawa}{\affiliation{Institute for Theoretical Physics,
Kanazawa University, Kanazawa 920-1192, Japan}}
\newcommand{\Moscow}{\affiliation{Institute of Theoretical and
Experimental Physics, B.Cheremushkinskaya 25, Moscow, 117218, Russia}}
\newcommand{\Pisa}{\affiliation{INFN,
Dipartimento di Fisica Universita di Pisa, Largo Pontecorvo 3, 56127 Pisa, Italy}}

\begin{document}

\title{Magnetic component of Yang-Mills plasma}

\author{M.N.~Chernodub}\Moscow\Kanazawa
\author{V.I.~Zakharov}\Pisa\Moscow

\begin{abstract}
Confinement in non-Abelian gauge theories is commonly ascribed to percolation of
magnetic monopoles, or strings in the vacuum. At the deconfinement phase
transition the condensed magnetic degrees of freedom are released into
gluon plasma as thermal magnetic monopoles. We point out that
within the percolation picture lattice simulations can be used to
estimate the monopole content of the gluon plasma. We show that
right above the critical temperature the monopole density remains a constant
function of temperature, as for a liquid, and then grows, like for a gas.
\end{abstract}


\date{November 29, 2006}

\maketitle

It is well known, the properties of the Yang-Mills plasma turned out to be unexpected.
In very brief, the plasma is similar rather to an ideal liquid than to
a gluon gas interacting perturbatively~\cite{shuryak}. Amusingly enough,
many features of the plasma find their theoretical explanation in terms
of a dual, or string  formulation of Yang-Mills theories~\cite{klebanov}
which has been derived only in the limit of infinite number
of colors and supersymmetry.

It is a challenge to uncover dynamical picture behind the observations on plasma.
In particular, equation of state has been studied in detail through lattice simulations
both in case of $SU(2)$ and $SU(3)$ gauge
theories~\cite{reviews}. It turns out that global characteristics of plasma are
not in contradiction either with perturbation theory or with the string picture,
or with quasiparticle models~\cite{quasi}, and do not provide much insight
into the plasma dynamics.

Infrared-sensitive  variables could be more helpful to identify specific degrees of
freedom of the plasma. An example of such a variable is the viscosity
which turns to be very low~\cite{teaney}. Other examples are the string
tensions $\sigma$ and $ \tilde{\sigma}$ of the spatial Wilson and 't Hooft loops,
\begin{equation}
\label{both}
\langle W_{\mathrm{spatial}}\rangle \sim \exp(-\sigma A)\,,\quad
\langle H_{\mathrm{spatial}}\rangle \sim \exp(-\tilde{\sigma} A)\,,
\end{equation}
respectively. Here $A$ is the area of a minimal surface spanned on the corresponding contour.

Thus, we come to consider light degrees of freedom of the plasma. On the theoretical side,
our guiding observation is that degrees of freedom condensed at $T=0$ form a light component of the thermal  plasma at $T>T_c$.
Since the confinement of color in non-Abelian theories is due to magnetic degrees of
freedom~\cite{thooft,greensite,digiacomo} the magnetic component is to be present in the plasma as well.

In the string picture one postulates existence of electric strings which can be open on external
quarks (Wilson line) and of magnetic  strings which can be open on external  monopoles ('t Hooft line).
At $T=0$, the electric strings have a non-zero tension, measured through
the Wilson line. One of the earliest proposals for the mechanisms of deconfinement
is percolation of electric strings \cite{polyakov}. Generically,
percolation means that there is large (potential) energy $E$ and large
entropy $S$ which cancel each other. Thus, within this mechanism it is long
strings which become tensionless and percolate.
Lattice data \cite{zantow} provides evidence for large entropy $S = - {(\partial F/\partial T)}_V$ of the electric strings,
since the average of the Wilson line $\langle W\rangle \sim \exp(-F/T)$
can directly be related to free energy~$F= E - T S$.
The potential energy continues to grow above the phase
transition, $T_c<T<2T_c$, and the entropy reaches  values of the order of 10, Ref.~\cite{zantow}.

On the lattice, the phenomenon of strong cancelation between energy and entropy
was discovered first for magnetic strings~\cite{ft}.
At zero temperature magnetic strings percolate through the vacuum
and provide disorder which causes confinement.
Percolating strings are directly observable on the lattice  and are known as center vortices~\cite{greensite}.
Non-Abelian action associated with the strings is ultraviolet--divergent,
$S_{\mathrm{strings}} \sim ({\mathrm{Area}})_{\mathrm{strings}}/a^2$,
where $a$ is the lattice spacing. The total area of the strings, on the other hand,
is a finite quantity in physical units,
$\langle ({\mathrm{Area}})_{\mathrm{strings}} \rangle \sim \Lambda_{\mathrm{QCD}}^2 V_{4d}$.
If there were no cancelation between the string energy and the string entropy, the area of the strings
would be in lattice units as well. Note that
there is no area law for the 't~Hooft loop at at $T=0$
and therefore the tension of the magnetic string vanishes.
Within large-$N_c$ dual formulations this relation is satisfied
explicitly~\cite{klebanov}.

Quantum mechanically, the lowest string mode is tachyonic.
The tachyonic mode corresponds to
the monopole condensation~\cite{digiacomo}.
Since there is only a single tachyonic mode
of the string, the string percolation can in fact be
projected into percolation of monopoles~\cite{ft}.

A link between percolation and field theory is
provided by the polymer formulation of field theory
in Euclidean space~\cite{ambjorn}.
One starts with classical action of a free particle,
$S_{\mathrm{cl}}=M \cdot L$, where $M = M(a)$ is a mass parameter
and $L$ is the length of trajectory. By evaluating the
Feynman path integral for the particle propagator one
learns that the propagating (physical) mass is in fact
\begin{equation}
\label{selfenergy}
m^2_{\mathrm{phys}}\,\sim\, \frac{C_0}{a}\big(M(a)
- \frac{C_{\mathrm{entr}}}{a}\big)\, ,
\end{equation}
where the constants are known explicitly for a particular (lattice, for example)
regularization. The tachyonic case corresponds, as usual,
to $m^2_{\mathrm{phys}}<0$.
The tachyonic mode is manifested as an infinite, or percolating
cluster. To ensure $m^2_{\mathrm{phys}} \sim \Lambda_{\mathrm{QCD}}^2$
the bare mass parameter $M(a)$ and the entropy
factor,  $C_{\mathrm{entr}}/a$ in Eq. (\ref{selfenergy}) are to be fine
tuned. This, crucial condition of fine tuning between energy and
entropy is satisfied for the lattice monopoles~\cite{ft}.
Moreover, percolation of the monopole at short distances
can be understood within the approximation~(\ref{selfenergy}),
while  at large distances the properties of the monopole
currents correspond to a constrained system \cite{maxim3}.
Observationally, the constraint is that the monopoles live on
two-dimensional surfaces, or strings, for a review see~\cite{vz5}.

It is a generic field-theoretic phenomenon that particles which are virtual at $T=0$,
are becoming real at finite temperature and released into the thermal plasma. In
our case, there exists a tachyonic monopole mode
which is to disappear at the point of the phase transition,
with magnetic strings emerging into the plasma~\cite{foot2}.
In order to realize a tempting opportunity to check this picture from the first principles
offered by Euclidean lattice simulations, one should be able to distinguish between real
and virtual particles in the Euclidean formulation of the theory. The problem is that
even real particles in the Euclidean space are off-mass-shell.

Let us first address the problem of detection of real particles in lattice
simulations in case of a free scalar field (for a related discussion
see Refs.~\cite{bornyakov,schakel}). Since the monopoles are observed as closed
trajectories on the lattice, it is appropriate to utilize
the polymer representation~\cite{ambjorn}. At zero temperature a
typical monopole ensemble consists of a large number of finite
clusters of the monopole trajectories corresponding to
virtual particles and one infinite cluster associated with
condensed (tachyonic) monopoles. As temperature increases
the infinite cluster disappears since the monopole condensate vanishes
at the deconfinement temperature. We point out that the monopole condensate
melts down into real thermal particles, which contribute to magnetic
content of the thermal plasma.

In the imaginary time formalism the finite temperature $T$ is imposed via compactification of the time
direction, $x_4$ into a circle of length $1/T$, and
the points $x = ({\mathbf{x}}, x_4 + s/T)$, $s \in \Z$,
are identified. In the
language of trajectories the integer number $s$ has
meaning of the wrapping number. It is obvious that properties of
thermal particles are encoded in the wrapped trajectories, $s \neq 0$, and
the virtual particles are non-wrapped, $s = 0$.

The propagator of a scalar particle is given by
\[
G(x - y) \propto \sum\nolimits_{P_{x,y}}e^{-S_{\mathrm{cl}}[P_{x,y}]} \,,
\]
where the sum is over all trajectories $P_{x,y}$ connecting points $x$ and $y$.
Evaluation of the propagator at a finite temperature $T$ is a straightforward
generalization of the $T=0$ case~\cite{ambjorn}. The propagator in momentum space,
\begin{equation}
{{\mathcal G}}_s({{\bf p}}) = \int \dd^3 {\mathbf x} \, e^{- i ({\mathbf p},{\mathbf x})} \, G({\mathbf x}, t = s/T)
\end{equation}
is related to the thermal distribution of the particles
\begin{equation}
\label{planck}
f_T(\omega_{\mathbf{p}}) = \frac{1}{2} \frac{{\mathcal G}^{\mathrm{wr}}
({\mathbf p})}{{\mathcal G}^{\mathrm{vac}}({\mathbf p})}\,,
\quad {\mathcal G}^{\mathrm{wr}} \equiv \sum_{s \neq 0} {{\mathcal G}}_s,
\quad {\mathcal G}^{\mathrm{vac}} \equiv {\mathcal G}_0,
\end{equation}
given by the Bose-Einstein formula $f_T = 1/(e^{\omega_{\mathbf{p}}/T}-1)$
where $\omega_{\mathbf{p}} = {({\mathbf{p}}^2 + m^2_{\mathrm{phys}})}^{1/2}$.

Equation~(\ref{planck}) demonstrates that wrapped trajectories in the Euclidean
space correspond to real particles in Minkowski space. However, the normalization
to the perturbative $T=0$ propagator, ${\mathcal G}^{\mathrm{vac}} = 4/ (a^2 \omega_{\mathbf{p}})$,
is awkward since it depends explicitly on the lattice spacing. Also, we
do not expect in fact that the
magnetic fluctuations of the Yang-Mills plasma -- as probed by the lattice monopoles --
correspond to free particles. As it is emphasized above, the monopoles represent only a
component of the whole plasma and their properties are to be constrained by the environment.

Moreover, one can explicitly show that the average number of
wrappings $s$ in a time slice  of volume $V_{3d}$ is most directly
related to the density of real particles
\begin{equation}
\label{exactt}
\rho(T) = n_{\mathrm{wr}} = \langle |s| \rangle/V_{3d}\,,
\end{equation}
where, in the case of the free particles,
\begin{equation}
\rho(T) = \int \frac{\dd^3 {\mathbf p}}{(2 \pi)^3} \frac{N_{\mathrm{d.f.}}}{e^{(\omega_{\mathbf{p}} + \mu)/T} - 1}\,,
\label{chemical}
\end{equation}
and $\mu$ is the (positively-defined) chemical potential controlling the particle number.
A similar relation is known for a gas of non-relativistic scalar particles~\cite{schakel}
and goes back to the Feynman's theory of $\lambda$-transition in ${}^4$He~\cite{feynman}.

The interaction of the monopoles with the environment is expressed, in particular,
in the number of effective degrees of freedom $N_{\mathrm{d.f.}}$ and in the non-zero chemical
potential $\mu$ to be discussed below. We will treat Eq.~\eq{exactt} as definition of the density
of thermal particles in the imaginary time formalism.

Thus, the density of the thermal particles $\rho(T)$ corresponds
to the vacuum expectation of the {\it number} of the wrapped loops.
By measuring the number of wrappings (in the Euclidean space) one
can learn density of real particles at finite temperature (in Minkowski space).

A change of the character of the monopole trajectories in Euclidean lattice simulations
near the point of the phase transition has been observed and discussed in
many papers, see, in particular, \cite{bornyakov,ejiri}. The change,
indeed, is disappearance of the percolating cluster and emergence of
wrapped trajectories. Quantitatively, only a part of the percolating (tachyonic)
cluster goes into the wrapped monopole trajectories, while the other part
is released into the vacuum as virtual particles.

We conclude that, qualitatively, there is little doubt that at $T=T_{c}$ there is
transition of a part of the tachyonic mode into degrees of freedom of the thermal plasma.
This is a spectacular phenomenon by itself and is a proof of reality of a magnetic
component of the gluon plasma.

To be quantitative, we need a detailed lattice data on the wrapped trajectories. In case
of pure $SU(2)$ gauge theory there is relevant data in literature, see, in particular,
Refs.~\cite{bornyakov,ejiri}. The most detailed data \cite{ejiri} refers, however,
rather to the length density than to the number density~\eq{exactt} of the wrapped trajectories
\begin{equation}
\label{ejiri}
\rho_{wr}\,\equiv\, L_{wr}/V_{4d}\,, \qquad V_{4d}\equiv V_{3d}/T\,,
\end{equation}
where $L_{wr}$ is the total length of the wrapped trajectories
in the volume $V_{4d}$. The expression for the density of the wrapping
number which enters the relation (\ref{exactt}) can be obtained
from (\ref{ejiri}) by replacing the total length of the wrapped
trajectories $L_{wr}$ by its projection on the time axis. The two
expressions coincide in the limit of static trajectories. In reality,
the approximation of static trajectories for the wrapped loops is
reasonable. Also, it has been checked \cite{ejiri} that the density
(\ref{ejiri}) is independent on the lattice spacing
and scales in the physical units, as is expected
for the density of the wrapping number.
Thus,  we use the data \cite{ejiri} on the wrapped trajectories
to estimate the density of real monopoles in plasma.

The data indicates existence of two distinct regions in the deconfinement phase:
the first region covers the range of temperatures $T_c < T < 2 T_c$, and the
second region corresponds to the higher temperatures $T > 2 T_c$.
In the first region the density of the thermal monopoles is almost insensitive to the
temperature~\cite{bornyakov,ejiri}
\begin{equation}
\label{density1}
\rho (T) \approx T_{c}^3 \qquad (T_c < T < 2 T_c)\,.
\end{equation}
The numerical value of the density -- corresponding, approximately, to 3.5 monopoles per cubic
fermi -- is comparable to the density of monopoles in the monopole condensate which amounts to
approximately $7.5$ monopoles per cubic fermi at zero temperature~\cite{Bornyakov:2005iy}.
Thus, a sizable fraction of the $T=0$ condensate is released at the transition region into the vacuum
as thermal particles.

It is instructive to compare the ratio~\eq{density1} with a limit of free relativistic particles.
Then the ratio~\eq{density1} at $T = T_c$ would be equal or smaller
than a well-known number, $\zeta(3)/\pi^2 \approx 0.12$. Thus, the monopole medium just above
the critical temperature is an order of magnitude denser than the ideal  gas estimate.
This fact along with the observation that the monopole density is independent on temperature
allows us to suggest that in the region $T_c < T < 2 T_c$ the monopoles form a dense (magnetic)
liquid.

There is another radical difference between the magnetic monopole
constituents of the Yang-Mills plasma and free particles.
For free particles, the wrapped trajectories are not static at all. Moreover,
the density of the wrapped trajectories diverges at
small lattice spacings $a$ as $\rho_{\mathrm{wr}}^{\mathrm{free}} \sim T^2/a$. On the other
hand, the available lattice data~\cite{ejiri} provides no indication of such a
divergence and this can be explained only by specific constraints, encoded, for example,
in chemical potential for the monopole trajectories, or
in the constraints that monopole trajectories belong to surfaces, see also \cite{maxim3}.

Starting from $T \approx 2T_c$ the density of particles associated with
the wrapped trajectories grows,
\begin{equation}
\label{density}
\rho (T) \approx (0.25 \, T)^3  \qquad (T > 2 T_c)\,,
\end{equation}
where we drop subleading terms in the asymptotic limit $T \to \infty$.
If we again compare~\eq{density} with the ideal gas case,
then in the asymptotic limit the monopoles take much less than one degree of freedom~\cite{foot3}.

According to the dimensional reduction arguments, valid at high temperature\cite{braaten}
non-perturbative physics is described by 3d magnetodynamics which corresponds
to zero Matsubara frequency of the original 4d theory.
In terms of the monopole trajectories the restriction to
the zero Matsubara frequency implies that the wrapped trajectories
become static. And indeed trajectories of 4d monopoles become static at higher temperature.
According to Refs.~\cite{suzuki:ejiri,hotgas} description of
monopoles in terms of 3d, or static trajectories and 4d wrapped
trajectories match each other at $T\approx 2.4 T_c$.

Matching with the dimensional reduction allows for an important consistency check of Eq.~\eq{exactt}.
Namely, as far as the monopole trajectories are static and, consequently, non-intersecting
they could be treated within a 3d theory~\cite{gas} utilizing methods of~\cite{polyakov2}.
Then the density~\eq{exactt} is equivalent to counting of monopoles trajectories.
This procedure is true also in case of strong interaction between the monopoles provided they are static.
The dimensional reduction implies that the monopole density~is
\begin{equation}\label{density2}
\rho(T) = C_\rho \, g^6_{3d}(T) \propto {\left(\frac{T}{\log T/\Lambda_{QCD}}\right)}^3 \qquad T \gg T_c\,,
\end{equation}
where $C_\rho$ is a temperature-independent parameter. Numerically, Eqs.~\eq{density} and \eq{density2} are
compatible with each other within accuracy of the available lattice data.

The temperature dependence exhibited by Eq (\ref{density2}) can be reproduced
by Eq. (\ref{exactt}) provided that there exists  temperature-dependent chemical
potential~\cite{foot3}
\begin{equation}
\mu \sim T \log g^{-6}_{4d}(T) \sim 3 T \log \log T / \Lambda\,,
\label{mu}
\end{equation}
which suppresses the monopole density~\eq{chemical} by the logarithmic
factor, $\exp\{- \mu /T\} \sim g^6_{4d}(T) \sim 1/ \log^3 (T/\Lambda)$.

Thus, the evolution of the magnetic component of the Yang-Mills vacuum
can schematically be represented as
\begin{equation}\label{chain}
\cd{\mbox{condensate}}{(T<T_c)}
\Longrightarrow
\cd{\mbox{magnetic liquid}}{(T_c < T < 2 T_c)}
\Longrightarrow
\cd{\mbox{gas}}{(T > 2 T_c)}
\end{equation}
In the confinement region the magnetic monopoles constitute a colorless tachyonic state
known as the monopole condensate while in the deconfinement state they form at first a
magnetized liquid and then a gas. In terms of the strings, percolation of both magnetic
and electric strings at $T>T_{cr}$ ensures area law for both
spatial  Wilson and 't Hooft loops, see (\ref{both}) \cite{shklovskii}.

Because of the limited accuracy of the lattice data existence of the chain (\ref{chain})
is established rather on qualitative level. Further numerical studies seem to
be well justified.

In conclusion, let us emphasize analogy between phenomena in Yang-Mills theory with physics of superfluidity.
In case of liquid helium there exist a superfluid and ordinary components of liquid.
With increasing temperature particles from the superfluid component are transferred to the ordinary-liquid
component~\cite{landau}. This is an analogy to the deconfinement phase transition with vacuum
condensate vanishing and magnetic degrees of freedom being released into the plasma around $T \approx T_c$.
In terms of the monopole trajectories the transition is from percolation in all directions to time-oriented
trajectories. The total density of percolating and wrapped trajectories remains approximately the same, in
analogy with a superfluid. The ``ordinary-magnetic-liquid'' component might be responsible for low viscosity of the plasma.

At higher temperature, $T>2T_c$
the density of the magnetic component grows and approaches perturbative regime, $\rho(T)\sim T^3$.
This is analog of evaporation of the liquid. One might expect that beginning with temperatures
$T\approx 2\,T_c$ the properties of plasma change gradually towards predictions
of perturbation theory.

\begin{acknowledgments}
The authors are supported by the JSPS grants No. L-06514, S-06032, and
thankful to A. Di Giacomo, F. V. Gubarev, J. Greensite, D. Kharzeev, A. Nakamura,
A. M. Polyakov, and T. Suzuki for useful discussions.
\end{acknowledgments}


\begin{thebibliography}{99}
\bibitem{foot3} {This can be interpreted as evaporation of the magnetic strings and, thus, monopoles into gluons at high~$T$.}
\end{thebibliography}
\end{document}